\documentclass[12pt]{article}
\begin{document}
\title{SOME CONSEQUENCES OF FUZZY SPACETIME}
\author{B.G. Sidharth\\
International Institute for Applicable Mathematics \& Information Sciences\\
Hyderabad (India) \& Udine (Italy)\\
B.M. Birla Science Centre, Adarsh Nagar, Hyderabad - 500 063 (India)}
\date{}
\maketitle
\begin{abstract}
In this paper we analyse the reasons which lead to a fuzzy spacetime approach. We then consider an experimental consequence, viz., a modified dispersion relation, which could be detected in ultra hligh energy cosmic rays, by for example NASA's GLAST Satellite scheduled to be launched in 2006. We also study the implications for the theory of non-Abelian Gauge Fields.
\end{abstract}
\section{Introduction}
There have been two approaches to fuzzy spacetime, both motivated by the ultra violet divergences encountered in Quantum Field Theory. The earliest approach was that of Snyder, Schild and others (Cf.refs.\cite{snyder,snyder2,schild,cu} and several references therein). These authors investigated a minimum space or time or spacetime interval or minimum cut off. This lead to what in modern terminology is called a Noncommutative Gometry:
$$[x,y] = (\imath a^2/\hbar )L_z,[t,x] = (\imath a^2/\hbar c)M_x,$$
\begin{equation}
[y,z] = (\imath a^2/\hbar ) L_x,[t,y] = (\imath a^2/\hbar c)M_y,\label{e1}
\end{equation}
$$[z,x] = (\imath a^2/\hbar )L_y,[t,z] = (\imath a^2/\hbar c)M_z,$$
The relations (\ref{e1}) are Lorentz invariant. The novelty of (\ref{e1}) is that if we retain terms $\sim (l^2)$ where $(l, \tau)$ represents the minimum cut off $a$, then the coordinates $x$ and $y$ no longer commute. This is the origin of the fuzzyness.\\
The other approach has been to endow points with a structure \cite{madore,madore2}. In this approach for example, a sphere is represented by, a point, or a pair of points (the North Pole or South Pole), or these poles together with the equator and so on depending on the degree of approximation. In other words the points on the sphere are ill defined.\\
In either case we end up with relations like
\begin{equation}
[x,y] = \Theta_{\imath j}0(xy)\label{e2}
\end{equation}
Relations like (\ref{e2}) once again show the fuzzyness - coordinates like $x$ and $y$ are represented typically by matrices and not points or real numbers.\\
It may be mentioned that this type of a geometry was originallyt envisaged by Dirac himself, in phase space. Starting from the equation for the position coordinate in the electron equation \cite{dirac}
\begin{equation}
x = (c^2p_1H^{-1}t) + \frac{\imath}{2} c\hbar (\alpha_1 - cp_1H^{-1}) H^{-1}\label{e3}
\end{equation}
Dirac argued that a literal interpretation of results as in (\ref{e3}) would imply that the electron has the speed of light and that the coordinates are complex, that is represented by non-Hermitian operators. So, he pointed out that in practise spacetime points are idealizations and really represent averages taken over the Compton scale. Once such zitterbewegung effects are eliminated by means of these averages, we are back effectively with physical spacetime. Infact this can be seen quite clearly from the theory of the energy-momentum (displacement) operators. In this case we have (Cf.ref.\cite{dirac})
\begin{equation}
\delta x (d_x + \bar d_x ) = \delta x^2 d_x\bar d_x = 0\label{e4}
\end{equation}
In (\ref{e4}) it is only when the squares of the minimum intervals are neglected that we get the usual four momentum operators - if these squared infinitessimals are retained then the operators become complex or non-Hermitian \cite{bgscsf}. So the momentum and hence the coordinate eigen values are complex in this latter case.\\
Once we realize the purport of the imaginary parts of the coordinates, in terms of averaging of the zitterbewegung effects and introducing minimum spacetime intervals, then we are back with the noncommutative geometry in (\ref{e1}) or (\ref{e2}).\\
Indeed directly starting from (\ref{e3}), the Dirac complex coordinate, it can be argued \cite{bgsfpl} that the noncommutative relations (\ref{e1}) or (\ref{e2}) follow, and moreover this can be modelled by a double Weiner process.\\ 
There is however an important nuance here. Neglecting squares of infinitessimal intervals implies, not that an area is being replaced by a point, but rather that it is being replaced by a length, viz., the minimum interval. That is why in this approximation, the missing dimension of the area shows up as an imaginary coordinate.\\
All this provides an explanation for the supposedly inexplicable reason why the Kerr-Newman metric arises from classical considerations with an imaginary shift of the coordinates \cite{newman,newman2,newman3}. This metric or classical considerations, correctly, gives the purely Quantum Mechanical $g=2$ factor of the electron. But the price that we have to pay is a singularity or a complex horizon:
\begin{equation}
r_+ = \frac{GM}{c^2} + \imath b,b \equiv \left(\frac{G^2Q^2}{c^8} + a^2 - \frac{G^2M^2}{c^4}\right)^{1/2}\label{e5}
\end{equation}
However it is remarkable that the electron coordinates given in the Quantum Mechanical Dirac formulation, (\ref{e3}) and the classical Newman formulation, (\ref{e5}) have the common feature that the imaginary parts are of the order of the Compton scale while the real parts represent the position of the electron. Indeed Newman himself was perplexed and observed \cite{newman4} ``... one does not understand why it (imaginary shift of origin) works. After many years of study I have come to the conclusion that it works simply by accident.''\\
The introduction of an imaginary coordinate, $x \to x+\imath x'$, can in the light of the above remarks be seen to be the neglect of an area or an extra or inner dimension. Indeed if we generalize to four dimensions then as is well known we get
\begin{equation}
(1, \imath ) \to (I, \vec \sigma )\label{e6}
\end{equation}
where $\sigma$'s represent the Pauli matrices (Cf. discussion in \cite{bgsfpl}). That is how an imaginary shift in a four dimensional context leads to the electron spin and the anomalous $g = 2$ factor, even in a classical theory. Coordinate shift given by (\ref{e6}) once again lead to a noncommutative geometry like (\ref{e1}) or (\ref{e2}).\\
We could argue that this is the generalization that was missed out in many complexification schemes, and infact in Nelson's stochastic theory, whlich conseqently did not lead anywhere (Cf.ref.\cite{bgsfpl}).\\
It may be pointed out that very much in the same spirit it was argued that we could think of inertial mass generation taking place due to what may be called self interaction within the Compton scale \cite{ijpap}. Within the Compton scale we have unphysical effects and a breakdown of speacial relativity. This problem was encountered nearly a century ago in the classical electron theory, where the electron self mass would tend to infinlity as its size shrunk to zero. On the other hand, classically the electron could not be given a finite size precisely because of unphysical effects due to the finite size and special relativity. The Quantum Mechanical Compton scale provides a cut off, shielding these supposedly unphysical effects \cite{bgsiced}. Indeed what we are doing here is something like a spacetime renormalization.\\
Finally it may be pointed out that \cite{moller} even in a purely classical theory of a collection of ultra relativistic particles we come agross a two dimensional disc of mass centres, with diameter of the order of the Compton scale, within whlich we encounter ``unphysical'' effects like negative energies.
\section{Some Consequences}
The fact that there is a minimum cut off that leads to a modification of special relativity and we have $(c=1=\hbar )$ a modified energy momentum relation \cite{bgsijtp2004}
\begin{equation}
E^2 = p^2 + m^2 - l^2 p^4\label{e7}
\end{equation}
Alternatively an equation like (\ref{e7}) would follow from a lattice formulation, but in this case, ultimately vanishing limits are taken \cite{lattice}.\\
Interestingly, it must be pointed out that modifications to Lorentz symmetry, such as (\ref{e7}) have been made on an ad hoc basis, \cite{cr,cr2,cr3,cr4} for the following reason:\\
If Lorentz symmetry were exactly true then ultra high energy cosmic rays, with energy somewhat greater than $10^{20}eV$ should not reach the earth. They would loose energy due to scatterling with the cosmic background radiation photons. This is the $GZK$ cut off (Cf. also ref.\cite{bgsijtp2004}). However, on the contrary some twenty such events have been observed, and it is possible that some or all of them could be attributed to a violation of Lorentz symmetry.\\
In any case returning to (\ref{e7}), if we consider particles with almost vanishing mass $m$, then we conclude that there is tachyonic behavior, as if the mass were imaginary, due to the presence of the extra term. This provides an explanation for the violation of the $GZK$ cut off alluded to above.\\
In any case the effect (\ref{e7}) leads to a modified dispersion relation as discussed in \cite{bgsijtp2004}, for example. Infact we have, a shift in frequency given by
$$\epsilon = \frac{l^2[Q^2 + 2mQ]^2}{2\{ m + k_0 (1 - cos \Theta )\}}$$
where $Q = k_0 - k$, the difference between incident and scattered frequencies.
It is quite possible that these effects pertaining to ultra high energy cosmic rays at energies which cannot be reproduced in accelerators will be detected by the GLAST Satellite to be launched by NASA in 2006.\\
Curiosly enough, if we were to take in (\ref{e7}), $l^2 < 0$, that is if we introduce an imaginary coordinate, then even if $m^2 \approx 0$, we can see that there is effectively a mass generation. This can immediately be explained by the fact alluded to above, that mass can be thought of as arising due to self interactions within the Compton scale, that is in the parlance of point spacetime, within this imaginary region.\\
Let us now consider the implications of not neglecting the second order coordinate differentials, within the context of the non-Abelian Gauge Theory. 
As is well known, this could be obtained as a generalization of the usual phase function $\lambda$ to include fields with internal degrees of freedom. For example $\lambda$ could be replaced by $A_\mu$ given by 
\begin{equation}
A_\mu = \sum_{\imath} A^\imath_\mu (x)L_\imath ,\label{e1a}
\end{equation}
The Gauge Field itself would be obtained by using Stoke's Theorem and (\ref{e1}). This is a very well known procedure: considering a circuit, which for simplicity we can take to be a parallellogram of side $dx$ and $dy$ in two dimensions, we can easily deduce the equation for the field, viz.,
\begin{equation}
F_{\mu \nu} = \partial_\mu A_\nu - \partial_\nu A_\mu - \imath q [A_\mu , A_\nu ],\label{e2a}
\end{equation}
$q$ being the Gauge Field coupling constant.\\
In (\ref{e2a}), the second term on the right side is typical of a non Abelian Gauge Field. In the case of the $U(1)$ electromagnetic field, this latter term vanishes.\\
Further as is well known, in a typical Lagrangian like 
\begin{equation}
\mathit{L} = \imath \bar \psi \gamma^\mu D_\mu \psi - \frac{1}{4} F^{\mu \nu} F_{\mu \nu} - m \bar \psi \psi\label{e3a}
\end{equation}
$D$ denoting the Gauge covariant derivative, there is no mass term for the field Bosons. Such a mass term in (\ref{e3a}) must have the form $m^2 A^\mu A_\mu$ which unfortunately is not Gauge invariant.\\
This was the shortcoming of the original Yang-Mills Gauge Theory: The Gauge Bosons would be massless and hence the need for a symmetry breaking, mass generating mechanism.\\
The well known remedy for the above situation has been to consider, in analogy with superconductivity theory, an extra phase of a self coherent system (Cf.ref.\cite{moriato} for a simple and elegant treatment). Thus instead of the Gauge Field $A_\mu$, we consider a new phase adjusted Gauge Field after the symmetry is broken
\begin{equation}
W_\mu = A_\mu - \frac{1}{q} \partial_\mu \phi\label{e4a}
\end{equation}
The field $W_\mu$ now generates the mass in a self consistent manner via a Higgs mechanism. Infact the kinetic energy term
\begin{equation}
\frac{1}{2} |D_\mu \phi |^2\quad ,\label{e5a}
\end{equation}
where $D_\mu$ in (\ref{e5a})denotes the Gauge covariant derivative, now becomes
\begin{equation}
|D_\mu \phi_0 |^2 = q^2|W_\mu |^2 |\phi_0 |^2 \, ,\label{e6a}
\end{equation}
Equation (\ref{e6a}) gives the mass in terms of the ground state $\phi_0$.\\
The whole point is as follows: The symmetry breaking of the gauge field manifests itself only at short length scales signifying the fact that the field is mediated by particles with large mass. Further the internal symmetry space of the gauge field is broken by an external constraint: the wave function has an intrinsic relative phase factor which is a different function of space time coordinates compared to the phase change necessitated by the minimum coupling requirement for a free particle with the gauge potential. This cannot be achieved for an ordinary point like particle, but a new type of a physical system, like the self coherent system of Superconductivity Theory now interacts with the gauge field. The second or extra term in (\ref{e4a}) is effectively an external field, though (\ref{e6a}) manifests itself only in a relatively small spatial interval. The $\phi$ or Higgs field in (\ref{e4a}), in analogy with the phase function of Cooper pairs of Superconductivity Theory comes with the Landau-Ginzburg potential $V (\phi)$.\\
Let us now consider in the Gauge Field transformation, an additional phase term, $f(x)$, this being a scalar. In the usual theory such a term can always be gauged away in the $U(1)$ electromagnetic group. However we now consider the new situation of a noncommutative geometry referred to above, 
\begin{equation}
\left[dx^\mu , dx^\nu \right] = \Theta^{\mu \nu} \beta , \beta \sim 0 (l^2)\label{e7a}
\end{equation}
where $l$ denotes the minimum spacetime cut off. (Cf. also ref.\cite{nc,annales,bgs}) (\ref{e7a}) is infact Lorentz covariant. Then the $f$ phase factor gives a contribution to the second order in coordinate differentials,
$$\frac{1}{2} \left[\partial_\mu B_\nu - \partial_\nu B_\mu \right] \left[dx^\mu , dx^\nu \right]$$
\begin{equation}
+ \frac{1}{2} \left[\partial_\mu B_\nu + \partial_\nu B_\mu \right] \left[dx^\mu dx^\nu + dx^\nu dx^\mu \right]\label{e8a}
\end{equation}
where $B_\mu \equiv \partial_\mu f$.\\
As can be seen from (\ref{e8a}) and (\ref{e7a}), the new contribution is in the term which contains the commutator of the coordinate differentials, and not in the symmetric second term. Effectively, remembering that $B_\mu$ arises from the scalar phase factor, and not from the non-Abelian Gauge Field, in equation (\ref{e2}) $A_\mu$ is replaced by 
\begin{equation}
A_\mu \to A_\mu + B_\mu = A_\mu + \partial_\mu f\label{e9a}
\end{equation}
Comparing (\ref{e9a}) with (\ref{e4a}) we can immediately see that the effect of noncommutativity is precisely that of providing a new symmetry breaking term to the Gauge Field, a term which does not come from the Gauge Field itself. Being an $0(l^2)$ effect, it manifests itself only at small scales, as required.\\
Effectively, because of (\ref{e9a}) we would have, specializing to a spherically symmetric field for simplicity, instead of the usual Maxwell equations in the gauge field context,
\begin{equation}
\vec E \to \vec E - \vec \nabla f = \vec \nabla Q - \vec \nabla f\label{e10a}
\end{equation}
So we have for a point Gauge charge, the modified equation
\begin{equation}
\nabla^2 Q = -4 \pi \rho + \lambda (r)\label{e11a}
\end{equation}
The solution of (\ref{e11a}) is
\begin{equation}
Q = \int_v \frac{(\rho + \lambda (r)}{r}\label{e12a}
\end{equation}
In (\ref{e11a}) and (\ref{e12a}) $\lambda (r)$ represents the effect of the noncommutativity and is an order of $l^2$ effect, that is it falls off rapidly. It can be seen that the first term in the integral on the right side of (\ref{e12a}) gives, in conjunction with (\ref{e10a}) the usual Coulumb type of a field. It is the second term in the integral which represents a field due to the noncommutativity of spacetime, which falls off rapidly, as it vanishes at scales where order of $l^2$ can be neglected. As such it represents a field mediated by massive particles.\\
(This is a well known example from the early days of Yang-Mills Theory, which lead to the conclusion that there was a Coulumb type potential of electromagnetism, that is a field without any mass.)\\
On the other hand if we neglect in (\ref{e7a}) terms $\sim l^2$, then there is no extra contribution coming from (\ref{e8a}) or (\ref{e9a}), so that we are in the usual non-Abelian Gauge Field theory, requiring a broken symmetry to obtain an equation like (\ref{e9a}). This is not surprising because if we neglect term $\sim l^2$ in (\ref{e7a}) then we are back with the usual commutative theory and the usual Quantum Mechanics.


\begin{thebibliography}{99}
\bibitem {snyder} H.S. Snyder, Physical Review, Vol.72, No.1, July 1 1947, p.68-71.
\bibitem {snyder2} H.S. Snyder, Physical Review, Vol.71, No.1, January 1 1947, p.38-41.
\bibitem {schild} A. Schild, Phys.Rev., 73, 1948, p.414-415.
\bibitem {cu} B.G. Sidharth, "Chaotic Universe: From the Planck to the Hubble Scale", Nova Science Publishers, Inc., New York, 2001.
\bibitem {madore} J. Madore, ``An Introduction to Non-Commutative Differential Geometry'', University Press, Cambridge, 1995.
\bibitem {madore2} J. Madore, ``Classical Quantum Gravity'', 9, 69-87, 1992.
\bibitem {dirac} P.A.M. Dirac, ``The Principles of Quantum Mechanics'', Clarendon Press, Oxford, 1958, p.263.
\bibitem {bgscsf} B.G. Sidharth, ``The Mystery of the Kerr Newman Metric'', to appear in Chaos, Solitons and Fractals.
\bibitem {bgsfpl} B.G. Sidharth, Foundation of Physics Letters, 16 (1), February 2003, 91-97.
\bibitem {newman} E.T. Newman, J.Math Phys., Vol.14, No.1, January 1973.
\bibitem {newman2} E.T. Newman and A.I. Janis, J.Math.Phys., Vol.6, No.6, June 1965, p.915-927.
\bibitem {newman3} E.T.  Newman, et al., J.Math.Phys., Vol.6, 1965, 918-919.
\bibitem {newman4} E.T. Newman, Enrico Fermi International School of Physics 1975, Proceedings p557.
\bibitem {ijpap} B.G. Sidharth, Ind.J.Pure and Appl.Phys., Vol.35, July 1997, pp.456-471.
\bibitem {bgsiced} B.G. Sidharth, ``An Interphase Between Classical Electrodynamics and Quantum Mechanics'' in ``Has the last word been said on Classical Electrodynamics'', Ed. A. Chubykalo et al., Rinton Press, USA, 2003.
\bibitem {moller} C. Moller, ``The Theory of Relativity'', Clarendon Press, Oxford, 1952, pp.170ff.
\bibitem {bgsijtp2004} B.G. Sidharth, ``Discrete Spacetime and Lorentz Symmetry'', International Journal of Theoretical Physics, 2004 (To appear).
\bibitem {lattice} I. Montway and G. Miinster, ``Quantum Fields on a Lattice'', Cambridge University Press, Cambridge, 1994, pp.164ff.
\bibitem {cr} S. Coleman and S.L. Glashow, PRD, 59, 116008, 1999.
\bibitem {cr2} T. Jacobson, xxx.astro-ph/0212190.
\bibitem {cr3} A.V. Olinto, Phys.Rev., 333-334, 2000, pp.329ff.
\bibitem {cr4} S.M. Carroll, Phys.Rev.Lett., 87, 2001, pp.141601ff.
\bibitem {moriato} K. Moriyasu, ``An Elementary Primer for Gauge Theory'', World Scientific, Singapore, 1983.
\bibitem {nc} B.G. Sidharth, Il Nuovo Cimento, 117B (6), 2002, 703ff.
\bibitem {annales}  B.G Sidharth, Annales de la Fondation Louis de Broglie, 27 (2), 2002, pp.333ff.
\bibitem {bgs} B.G. Sidharth in ``Proceedings of the Fifth International Symposium on Frontiers of Fundamental Physics'', Universities Press (Orient Longman), 2004 (in press).
\end{thebibliography}
\end{document}